%Paper: hep-th/9511164
%From: OOGURI@theor3.lbl.gov
%Date: Wed, 22 Nov 1995 16:57:35 -0800 (PST)
%Date (revised): Wed, 22 Nov 1995 17:50:20 -0800 (PST)

\input harvmac
\Title{\vbox{\hbox{HUTP-95/A045, LBL-37996, UCB-PTH-95/41}
\hbox{\tt hep-th/9511164}}}
{Two-Dimensional Black Hole and Singularities of CY Manifolds}
\bigskip
\centerline{Hirosi Ooguri}
\bigskip\centerline{\it Department of Physics, University of
California at Berkeley}
\centerline{\it 366 Le\thinspace Conte Hall, Berkeley, CA 94720-7300}
\centerline{and}
\centerline{\it Theoretical Physics Group, Mail Stop 50A-3115}
\centerline{\it Lawrence Berkeley  Laboratory,
                Berkeley, California 94720}

\vskip .1in
\centerline{and}

\vskip .1in

\centerline{Cumrun Vafa}
\bigskip\centerline{\it Lyman Laboratory of Physics}
\centerline{\it Harvard University}\centerline{\it Cambridge, MA 02138}

\vskip .3in
We study the degenerating limits of superconformal
theories for compactifications on singular $K3$ and Calabi-Yau threefolds.
We find that in both cases
the degeneration involves creating an Euclidean two-dimensional
black hole coupled weakly to the rest of the system.
Moreover we find that the conformal theory of
$A_n$ singularities of $K3$ are the same as that of the symmetric
fivebrane.  We also find intriguing connections
between $ADE$ $(1,n)$ non-critical strings
and singular limits of  superconformal theories
on the corresponding ALE space.
\Date{\it {November 1995}}
%\draft

\vskip 1in

\centerline{\bf Disclaimer}

\vskip .2in

This document was prepared as an account of work sponsored by the United
States Government. While this document is believed to contain correct
information, neither the United States Government nor any agency
thereof, nor The Regents of the University of California, nor any of their
employees, makes any warranty, express or implied, or assumes any legal
liability or responsibility for the accuracy, completeness, or usefulness
of any information, apparatus, product, or process disclosed, or represents
that its use would not infringe privately owned rights.  Reference herein
to any specific commercial products process, or service by its trade name,
trademark, manufacturer, or otherwise, does not necessarily constitute or
imply its endorsement, recommendation, or favoring by the United States
Government or any agency thereof, or The Regents of the University of
California.  The views and opinions of authors expressed herein do not
necessarily state or reflect those of the United States Government or any
agency thereof, or The Regents of the University of California.

\vskip 2in

\centerline{\it Lawrence
Berkeley  Laboratory is an equal opportunity employer.}

\vfill

\newsec{Introduction}
One of the central issues in the recent progress in understanding
non-perturbative aspects of string theory has been the
realization that singular conformal theories which correspond
to singular points in moduli of string compactifications
in many cases signal the existence of nearly massless
solitons \ref\hut{C.M.\ Hull and P.K.\ Townsend,
``Unity of superstring dualities,'' {\it Nucl. Phys.} {\bf B438}
(1995) 109.},
\ref\wittenduality{E.\ Witten, ``
String theory dynamics in various dimensions,''
{\it Nucl. Phys.} {\bf B443} (1995) 85.},
\ref\strom{A.\ Strominger,
``Massless
black holes and conifolds
in string theory'', {\it Nucl. Phys.} B451 (1995) 96.},
\ref\pol{J.\ Polchinski, ``Dirichlet branes and Ramond-Ramond charges,''
e-Print Archive: hep-th/9510017.}.
Since these nearly massless modes are left out of the low energy spectrum of
elementary excitations of strings, integrating them out will
lead to a singular low energy description.  As examples of
this phenomenon one can mention the enhanced gauge symmetry
of type IIA once we compactify on singular limits of $K3$ (or ALE
spaces)
where the gauge group is $ADE$ depending
on the singularity type \wittenduality\ and the Dynkin diagram
is determined by the intersection form of vanishing
2-cycles.  Another example is provided by the
compactifications of type IIB strings on singular
Calabi-Yau threefolds.  If we have a conifold singularity,
where a 3-cycle vanishes, one expects a massless solitonic
hypermultiplet (carrying Ramond-Ramond charge)\strom.
It is therefore important to understand the nature of
the singularity of the conformal theory in these classes
of examples.  This is the main aim of this paper.
As a byproduct, the results presented here have led
to a dual perturbative string description of the massless solitons
in both of these classes \ref\dstring{
M.\ Bershadsky, C.\ Vafa and V.\ Sadov,
``D-strings and D-manifolds,'' e-Print Archive: hep-th/9510225.},
where the nearly massless soliton is realized as an elementary
state of the  D-string.

The organization of this paper is as follows.  We will
first discuss the case of Calabi-Yau 3-fold in the
presence of conifold singularity.  Then we consider
ALE spaces with $ADE$ singularities and present
a conformal theory description of them.  Rather
surprisingly we find that they are equivalent to
symmetric fivebrane solutions.  We will establish
this at the level of conformal theories and give a
geometric explanation of this phenomenon
based on `stringy cosmic strings' \ref\scs{
B.R.\ Greene, A.\ Shapere, C.\ Vafa and S.-T.\ Yau,
``Stringy cosmic strings and non-compact Calabi-Yau manifolds,''
{\it Nucl. Phys.} {\bf B337} (1990) 1.}.
 Also some mysterious
connections are found with the $(1,n)$ conformal theories
coupled to gravity, following a similar connection between
the $c=1$ at the self-dual radius and the conifold singularity
of Calabi-Yau threefold \ref\gv{D.\ Ghoshal and C.\ Vafa, ``$c=1$ string
as the topological theory of the conifold,'' e-Print Archive:
hep-th/9506122.}.  In particular
just as the conifold singularity is topologically
captured by the $N=2$ topological theory,
one expects that the $ADE$ singularities
on ALE are captured by $N=4$ topological theory
\ref\bv{ N.\ Berkovits and C.\ Vafa,
``$N=4$ Topological
Strings,'' {\it Nucl. Phys.} {\bf B433} (1995) 123.}.  We
find indications that this latter $N=4$ topological system is
related and possibly equivalent to $(1,n)$ conformal system coupled to
the two-dimensional gravity.

\newsec{Conifold in Three Dimensions and Massless Soliton}

We can gain information on the spectrum of the conformal field theory
by computing \ref\bcovone{M.\ Bershadsky, S.\ Cecotti, H.\ Ooguri and C.\ Vafa,
``Holomorphic Anomalies in Topological Field Theories,''
{\it Nucl. Phys.} B405 (1993) 279.}
$$ F_1 = {1 \over 2} \int_{{\cal M}_1}
  {d^2 \tau \over {\rm Im} \tau} {\rm tr}' (-1)^{F_L + F_R}
  F_L F_R q^{H_L} \bar{q}^{H_R} , $$
where ${\rm tr}'$ means a trace over the Hilbert space of
the $N=2$ superconformal non-linear sigma-model on the Calabi-Yau
threefold in the Ramond sector with zero-energy (chiral) states eliminated,
and $F_L$ and $F_R$ are left and right-moving fermion number operators.
This is a standard expression for one-loop partition function
of a topological string on the Calabi-Yau manifold, but it can also be
interpreted as computing one-loop correction to $R^2$ terms in the
low energy effective action of the type II superstring
theory \ref\bcovtwo{M.\ Bershadsky, S.\ Cecotti, H.\ Ooguri
and C.\ Vafa, ``Kodaira-Spencer theory of gravity and exact results
for quantum string amplitudes'', {\it Commun. Math. Phys.} {\bf 165} (1994)
311.}, \ref\narain1{ I.\ Antoniadis, E.\ Gava, K.S.\ Narain and T.R.\ Taylor,
``Topological amplitudes in string theory,''
{\it  Nucl. Phys.} {\bf B413} (1994) 162.}. In general,
$F_1$ computed for a compact Calabi-Yau manifold is finite
since the spectrum of the sigma-model should be discrete
with a finite mass gap and therefore  $ {\rm tr}' (-1)^{F_L + F_R}
F_L F_R q^{H_L} \bar{q}^{H_R}$ becomes exponentially small
as $\tau \rightarrow \infty$.  In \bcovone, $F_1$ is
computed in four different families Calabi-Yau manifolds, each of which
has a conifold singularity in the moduli space.
In the neighborhood of the conifold, local structure
near the singularity of the Calabi-Yau manifolds can be described
in the universal form as the quadric \ref\conifold{ P.\ Candelas, A.M.\ Dale,
C.A.\ L\"utken, and R.\ Schimmrigk,
``Complete Intersection Calabi-Yau Manifolds'', {\it Nucl. Phys.} {\bf
B298} (1988)
493;  P.S.\ Green and T.\  H\"ubsch,
``Possible Phase Transitions Among Calabi-Yau Compactifications'',
{\it Phys. Rev. Lett.} 61 (1988) 1163;
``Connecting Moduli Spaces of Calabi-Yau Threefolds'', {\it Commun. Math.
Phys.} 119 (1988) 431;
P.\ Candelas, P.S.\ Green, and T.\
H\"ubsch,
``Finite Distance Between Distinct Calabi-Yau Manifolds'',
{\it Phys. Rev. Lett.} 62 (1989) 1956;
``Rolling Among Calabi-Yau Vacua'', Nucl. Phys.
 B330 (1990) 49;
 P.\ Candelas and X.C.\ de la Ossa,
``Comments on Conifolds'', {\it Nucl. Phys.} {\bf B342} (1990) 246.}
\eqn\quadric{ y_1^2 + y_2^2 + y_3^2 + y_4^2 = \mu, ~~(y \in {\bf C}^4) .}
As $\mu \rightarrow 0$, the quadric develops a singularity with
a collapsing $3$-cycle. It turned out that in all the four cases,
$F_1$ behaves as \foot{At genus-one, the topological string actually
computes $\partial_\mu F_1$ since we need to fix
one-point on a torus.  More generally for $F_g$ we would
be interested in $F_{g | \bar{\mu} \rightarrow 0}$ which is the
topological expansion of $F_g$ in the conifold limit.}
\eqn\oneloop{  F_1 \sim - {1 \over 12} \log |\mu|^2   }
as $\mu \rightarrow 0$.  This was further confirmed
by computation of $F_1$ in many other examples studied
in \ref\other{P.\ Candelas,
A.\ Font, S.\ Katz and D.\ Morrison, ``Mirror symmetry
for two parameter models 1'' {\it Nucl. Phys.}
{\bf B416} (1994) 481; ``2," {\it Nucl. Phys.}
{\bf B429} (1994) 626; S.\ Hosono, A.\ Klemm, S.\ Theisen
and S.-T.\ Yau, ``Mirror symmetry, mirror map and applications
to complete intersection Calabi-Yau spaces,''
{\it Nucl. Phys.} {\bf B433} (1995) 501.}.
To be consistent with Strominger's interpretation,
this logarithmic divergence
should be due to the virtual nearly massless solitonic loop \ref\vafaone{C.\
Vafa, ``A stringy test of
the fate of the conifold'',  {\it Nucl. Phys.} {\bf B447} (1995) 252.}.
On the other hand, from the point of view of conformal field theory, this
divergence signals vanishing of mass gap in the spectrum since contributions
from massive excitations to $F_1$ are finite. This implies that the
perturbative string theory acquires additional light particles in the
neighborhood of the conifold. These light particles are
unstable in string perturbation theory.

Hints that the topological theory of Calabi-Yau threefold are related
to $c=1$ strings at self-dual
radius first appeared in \ref\mv{S.\ Mukhi and C.\ Vafa,
``Two-dimensional black-hole as a topological coset model of
$c=1$ string theory,'' {\it Nucl. Phys} {\bf B371} (1992)
191.}\ where it was shown to be described by a $\hat c=3$
twisted Kazama-Suzuki model \ref\ks{
Y.\ Kazama and H.\ Suzuki, ``New $N=2$ superconformal field theories
and superstring compactification,'' {\it Nucl. Phys.}
{\bf B321} (1989) 232.} for a coset $SL(2)/U(1)$ at level $k=3$.
In fact this explained the observation in \ref\witN{E.\ Witten, ``The N matrix
model and gauged WZW models'', {\it Nucl.\ Phys.}{\bf B371} (1992) 191.}\
that the topological theory of this model at genus-$g$
gives Euler characteristic of moduli space of Riemann surfaces
which is the same as the partition function of $c=1$ string
at the self-dual radius \ref\matrix{I.\ Klebanov and D.\ Gross,``
One-dimensional string theory on a circle'', {\it Nucl. Phys.}
{\bf B344} (1990) 475.}, \ref\dv{J.\ Distler and C.\ Vafa,
``A critical matrix model at $c=1$'', {\it Mod. Phys. Lett.} {\bf A6}
(1991) 259.}.
The equivalent Landau-Ginzburg description for it
is given by a superpotential $W=-\mu x^{-1}$ by
``analytic continuation'' from the $SU(2)/U(1)$ Kazama-Suzuki model
where a superpotential is $W=x^{k+2}$ for level $k$ \ref\m{E.J.\
Martinec, ``Algebraic geometry and effective Lagrangians,''
{\it Phys. Lett.} {\bf 217B} (1989) 431.}, \ref\vw{C.\ Vafa and N.\ Warner,
``Catastrophes and the classification of conformal theories,''
{\it Phys. Lett.} {\bf 218B} (1989) 51.}. More evidence in
favor of this description was found by studying
the correlators of this model in the Landau-Ginzburg
framework and obtaining the $c=1$ tachyon amplitudes
\ref\gmtwo{D.\ Ghoshal and S.\ Mukhi, ``Topological
Landau-Ginzburg model for two-dimensional string theory'', {\it
Nucl.\ Phys.}{\bf B425} (1994) 173.}, \ref\hop{A.\ Hanany, Y.\ Oz and
R.\ Plesser,``Topological Landau-Ginzburg formulation and integrable
structure of 2-D string theory'',{\it Nucl. Phys.} {\bf B425} (1994) 150.}.

Further evidence for the  $c=1$ -- Calabi-Yau connection
emerged from \bcovtwo\ where it was
noticed that as a function of $\mu$ the genus-$g$ partition
function of topological theory near conifold limit
scales as $F_g\sim \mu^{2-2g}$,
which is the same as one expects for $c=1$ theory coupled to gravity.
This connection was explained
in \gv\  by
noting that the sigma-model for \quadric\ is related to the
Landau-Ginzburg model with a superpotential
$$ W(x,y) = - \mu x^{-1} + y_1^2 + y_2^2 + y_3^2 + y_4^2 .$$
The equation $W=0$ in a weighted projective space
$WCP^4_{-2,1,1,1,1}$ satisfies the Calabi-Yau condition with
the central charge $\hat{c} = 3$ \ref\gmone{D.\ Ghoshal and S.\ Mukhi,
``Landau-Ginzburg model for a critical topological string,''
to appear in the Proceedings of the
International Colloquium of Modern Quantum Field Theory, eds S. Das et al.}.
Note that adding quadratic fields does not change the conformal
fixed point and is a
standard technique for relating Calabi-Yau compactifications
with Landau-Ginzburg theories
\ref\vgw{B.R.\ Greene, C.\ Vafa and N.P.\ Warner,
``Calabi-Yau manifolds and renormalization group flow,''
{\it Nucl. Phys.} {\bf B324} (1989) 371.}.
When $x \neq 0$, we can set $x=1$ without any loss of generality
and $W=0$ in this patch is equivalent to the deformed conifold \quadric.
As far as $\mu \neq 0$, the patch with $x=0$ is associated with
$\partial W = \infty$ and would not arise in the geometric description
of the theory
(at least in a suitable
regime of field space -- it would be interesting
to check these from the viewpoint of linear sigma models
 \ref\witphase{E.\ Witten, ``Phases of $N=2$
theories in two dimensions'', {\it Nucl.\ Phys.}{\bf B403} (1993) 159.}).
These results in particular explain the universal behavior
of $F_1$ near the conifold singularity.  Moreover
for $g>1$, using the result for $c=1$ string partition function
we learn that $F_g$ of the Calabi-Yau near the conifold limit
is given by
\eqn\penner{  F_g = {B_{2g} \over 2g (2g-2)} \mu^{2-2g} , }
where $B_n$ is the Bernoulli number and $B_{2g}/ 2g (2g-2)$ is
the Euler class of the moduli space of genus-$g$ Riemann surfaces.
At $g=2$, this expression is consistent with the conifold limit of
the result in \bcovtwo\ where the $g=2$ topological string amplitude
is computed for the quintic threefold. More recently
it was shown that the behavior \penner\ is consistent to all
order in $g$ with the corresponding correction to $R^2 F^{2g-2}$
in four dimensions where one considers a virtual loop of a nearly
massless soliton \ref\agnt{I.\ Antoniadis, E.\ Gava, K.S.\ Narain
and T.R.\ Taylor, ``$N=2$ type II -- heterotic duality and higher
derivative $F$-terms,'' e-Print Archive:  hep-th/9507115.}\
thus generalizing the work in \vafaone\ to all loops and
providing a strong check for the conjecture of Strominger that
a nearly massless solitonic object `explains' topological singularity
of conformal theory in the conifold limit.

\subsec{Geometrical Picture of the Conformal Theory}
The conformal theory describing the conifold is thus the untwisted
version of the $SL(2)/U(1)$ Kazama-Suzuki model
with $k=3$. The non-supersymmetric version of this is
the Euclidean black hole system in two dimensions \ref\witb{E.\ Witten,
``On string theory and black holes,'' {\it Phys. Rev.} {\bf D44}
(1991) 314.}.
The geometrical picture developed there is as follows (see also
\ref\dvv{R.\ Dijkgraaf, H.\ Verlinde and
E.\ Verlinde,`` String propagation in a black hole geometry,''
{\it Nucl. Phys.} {\bf B371} (1992) 269.}):
The geometry is that of an infinitely
long ``cigar.''  For most of the geometry, if
we are far enough from the tip of the cigar, we have a cylindrical
geometry parameterized by $X$ running over a circle and $\phi$
along the length of cigar.   Moreover there is a linear dilaton in the
$\phi$ direction with the string coupling becoming weaker as we go along
the cigar.   The supersymmetric description has basically the same
bosonic piece, but in addition the standard fermionic partners.
As for the physical states of the $c=1$ string at the self-dual radius,
they come from the discrete representations of $SL(2)$ \mv\ and they can
be viewed as bound states at zero energy near the tip of the
cigar \dvv .  On the other hand the waves propagating down
the cigar correspond to non-topological states, and are BRST
trivial in the sense of the $c=1$ string -- However those are precisely
the ones which are responsible for the continuous spectrum of the
Calabi-Yau near the conifold.  This interpretation is consistent
with the considerations of unitarity:  We expect that the conformal
theory of Calabi-Yau has a unitary spectrum.  The
analysis of \ref\di{L.J.\ Dixon, M.E.\ Peskin and J.\ Lykken,
``$N=2$ superconformal symmetry and $SO(2,1)$ current
algebra, {\it Nucl. Phys.} {\bf B325} (1989) 329.}\
concluded that for this level of $k$ essentially none
of the discrete representations of the coset model is unitary
(the exception being the one corresponding to the cosmological constant
operator which is also a real deformation of Calabi-Yau), whereas
all the continuous representations are unitary.   In fact this meshes
nicely with the picture advocated in
\ref\witju{E.\ Witten, ``Some comments on string dynamics,''
e-Print Archive:  hep-th/9507121.}\ if we
think of the length of the tube as being related to $-{\rm log}\mu$
and connected to the rest of the Calabi-Yau conformal theory.
The states of $c=1$ string which are bound to the tip of the cigar
do not communicate with the rest of the Calabi-Yau because the coupling
is weaker by a factor of $\mu$,
and moreover they
do not arise as physical states of the Calabi-Yau compactification,
consistent with unitarity requirements.

The picture advocated in \witju\ was in connection with the explanation
of the unreliability of string perturbation theory.  In such a
picture the coupling being infinitely strong at the tip of the cigar
in the limit $\mu \rightarrow 0$ shows that the perturbation theory
should break down if $\mu < \lambda$, where $\lambda$ is the string
coupling constant.   In fact this already follows
from the computation of the topological amplitude above:  Namely
if we turn on a background $F$ and compute the correction to $R^2$,
as explained above,
near the conifold the genus-$g$ correction goes as
$F_g\sim \mu^{2-2g}\lambda^{2g-2}$ and thus if $\mu <\lambda$ each
successive correction gets larger and the perturbation sum
becomes unreliable\foot{This story is very similar
and possibly related to the issue of heterotic/type I duality
recently discussed in \ref\polw{J.\ Polchinski and E.\ Witten,
``Evidence for heterotic -- type I string duality,''
e-Print Archive: hep-th/9510169.}.}.

The main puzzle is how does this fit with the geometry of
the Calabi-Yau near the conifold?  To begin with we have a vanishing $S^3$.
Where did the cigar shaped object appear?  When did we turn on the dilaton?
We will be able to shed some light on this after our discussion
on the singularities of the conformal theory near $K3$ degenerations.

\newsec{$ADE$ Singularities in $K3$ and Fivebrane Solution}

Let us turn our attention to singularities in $K3$ surfaces.
{}From the duality between type II and heterotic string models
in six dimensions, it is expected that the perturbative type II string
on $K3 \times {\bf R}^6$ will have singularities when the dual heterotic
string
on $T^4 \times {\bf R}^6$ approaches enhanced gauge symmetry
points \wittenduality.
These special points for $K3$ are modeled
(and generalized) by the ALE spaces which have
$ADE$-type singularities
given by
\eqn\ade{
\eqalign{ A_{n-1} : & ~~ y_1^n + y_2^2 + y_3^2 = \mu ~~(n \geq 2)\cr
          D_{{n \over 2}+1} : & ~~ y_1^{{n \over 2}} + y_1 y_2^2 + y_3^2 = \mu
   ~~(n:{\rm even} \geq 6) \cr
          E_6 : & ~~ y_1^4 +y_2^3 + y_3^2 = \mu \cr
          E_7 : & ~~ y_1^3 y_2 +y_2^3 + y_3^2 = \mu \cr
          E_8 : & ~~ y_1^5 +y_2^3 + y_3^2 = \mu, ~~~(x \in {\bf C}^3) \cr } }
with appropriately chosen values of the K\"ahler moduli
\ref\aspin{P.S.\ Aspinwall, ``Enhanced gauge symmetries and $K3$
surfaces,'' {\it Phys. Lett.} {\bf 357B} (1995) 329.}.
As in the case of Calabi-Yau threefolds, we can add one more variable
$x$ and consider corresponding Landau-Ginzburg models with
superpotentials
\eqn\adesuperpotential{
\eqalign{ A_{n-1} : & ~~ W = -\mu x^{-n} + y_1^n + y_2^2 + y_3^2  \cr
          D_{{n\over 2}+1} : & ~~ W = -\mu x^{-n} + y_1^{{n \over 2}}
                    + y_1 y_2^2 + y_3^2  \cr
          E_6 : & ~~ W = -\mu x^{-12} + y_1^4 +y_2^3 + y_3^2  \cr
          E_7 : & ~~ W = -\mu x^{-18} + y_1^3 y_2 +y_2^3 + y_3^2  \cr
          E_8 : & ~~ W = -\mu x^{-30} + y_1^5 +y_2^3 + y_3^2 . \cr } }
It is straightforward to verify that the equation $W=0$ satisfies
the Calabi-Yau condition with $\hat{c} = 2$.
Note also that,
 as is usual in Gepner
models, to get a connection with Calabi-Yau, we have to mod out
by ${\rm exp}(2\pi iJ_0)$ which imposes the charge integrality
condition and summing over the spectral flow \ref\gepner{D.\
Gepner, ``Exactly solvable string compactification on manifolds of
$SU(N)$ holonomy,'' {\it Phys. Lett.} {\bf 199B} (1987) 380.}.
For example for the $A_{n-1}$ series this results in modding out
by a $Z_n$.

Let us analyse the Hilbert space structure of these models.
The $y$-dependent parts of \adesuperpotential\ are nothing
but the superpotentials of the $N=2$ superconformal minimal models
associated to the $ADE$-type modular invariants \m , \vw\
 with
$$ \eqalign{
\hat{c} = 1-{2 \over n}
{}~~~~~(&{\rm for}~~A_{n-1}~~{\rm and}~~D_{{n\over 2}+1} \cr
                      & n=12, 18, 30 ~~{\rm for}~~ E_6, E_7, E_8). \cr}$$
On the other hand, the Laudau-Ginzburg model with a potential $x^{-n}$
is equivalent to the Kazama-Suzuki model for the coset
$SL(2,R)_{n+2}/U(1)$ at the critical point.
Above we presented evidence for this analytic continuation in the
form of the superpotential $W$
from $SU(2)$ to $SL(2)$ when $n=1$ and we are assuming this is
generally valid.
 Therefore the
Landau-Ginzburg models given by the potentials \adesuperpotential\
are products of the Kazama-Suzuki model and the $N=2$ minimal models
modded out by an appropriate discrete group:
\eqn\geome{{{SL(2)\over U(1)}\times {SU(2)\over U(1)}\over G}}
where $G=Z_n$ in the $A_{n-1}$ case.
Before going to the detail of the conformal theory
construction and in particular the orbifoldizing of \geome\ by $G$,
we would like to mention what the outcome is.  Let us first
consider the $A_{n-1}$ series.
The bosonic piece of $SL(2)/U(1)$ coset model
can be viewed geometrically as a cigar shaped object \witb\
which is physically the Euclidean two-dimensional black hole.  This
is a non-compact space, and except for a finite region near the tip of the
cigar it can be viewed as a semi-infinite cylinder.  We will coordinatize
the length of it by $\phi$ and the circle coordinate we will denote
by $\tilde X$ with radius $R=\sqrt{2n}$ (which can be read off
from the level of the affine Lie algebra).  Moreover, there is a linear
dilaton for $\phi$, i.e. it is a Feigin-Fuchs system (this is
true if we are far enough from the tip of the cigar).
In addition we have, in this region, two free fermionic partners
of $\phi$ and $\tilde X$.
  As for the
supercoset $SU(2)/U(1)$, it is more difficult to describe geometrically,
but basically they are the representation of the parafermionic algebra
together with a free boson (whose momentum is
correlated with representations  of the parafermionic system).   In modding
out by $G= Z_n$,  the $Z_n$ acts trivially on the fermionic pieces
of $SL(2)/U(1)$ coset model,
but acts on the momentum (winding) states around $\tilde X$ by $n$-th
roots of unity.  At the same time, it also acts on the free bosonic part of the
$SU(2)/U(1)$ minimal model.  The net effect of this modding is to
recombine  $\tilde X$ and the free boson in the minimal model into
two bosons of radii $\sqrt{2(n-2)}$ and 1. The momentum of the
boson with radius $\sqrt{2(n-2)}$, which we denote by $H$,
is still correlated to the representations of the parafermion algebra.
On the other hand, the boson at $R=1$ is decoupled and can be
converted into two free fermions.  The parafermionic
system together with the $H$ field forms a bosonic $SU(2)$ system.
 Thus we finally end up after orbifoldization by $G$ with
a bosonic system consisting of $SU(2)$ at level $k=n-2$ and
a Feigin Fuchs system $\phi$ with background charge.
We have also 4 free fermions, three of
which make an $N=1$ superconformal $SU(2)$ model and the last one
is a superpartner of $\phi$.
 The background charge  of $\phi$ is tuned
so that the total ${\hat c}$ equals $2$. This is the same as the symmetric
fivebrane conformal theory \ref\chs{C.G.\ Callan, J.A.\ Harvey and
A.\ Strominger, ``Worldsheet approach to heterotic solitons and
instantons,'' {\it Nucl. Phys.} {\bf B359} (1991) 611.}\
with a charge associate to the $B$ field
($H$ charge) equal to $n$.
Note that our system is a {\it capped} version of
 \chs , in the
sense that instead of the $\phi$ going off to infinity, where
the coupling becomes infinitely strong, we end up instead at
the tip of the cigar.  This is the region where the splitting
of the $SL(2)/U(1)$ coset system to $\phi,\tilde X$ fails.

Now we come to a more detailed description of the orbifoldization
above using the characters of conformal field theory, concentrating
on how the orbifoldization by $G$ leads to the $SU(2)$ characters.
To perform the orbifoldization
\geome , we first note that the $U(1)$ current $J$ of the
$N=2$ algebra in the Kazama-Suzuki model $SL(2,R)_{n+2}/U(1)$ takes the
form
\eqn\uone{ J=\psi \bar{\psi}+ i \sqrt{{2 \over n}} \partial X }
where $\psi$, $\bar{\psi}$ are free fermions that are superpartners
of the bosonic coset $SL(2,R)/U(1)$ and $X$ is a free scalar living
on $S^1$ with radius $R = \sqrt{2/n}$ (where $X$ is dual to $\tilde
X$ described above).
 This can be shown in the following
way. In the Kazama-Suzuki model, the $N=2$ superconformal algebra are
realized as
\eqn\kazama{ \eqalign{ G^+ & = \psi J^+, ~~G^- = \bar{\psi} J^- \cr
              J & = (1 +{2\over n}) \psi \bar{\psi} - {2 \over n} J^3 \cr} }
where $J^\pm, J^3$ are the generators of $SL(2,R)_{n+2}$. To take the
quotient by $U(1)$, we introduce an $U(1)$ gauge field $A = i \partial X$,
BRST ghosts $(B,C)$ of weights $(1,0)$ and a BRST current
$$ J_{BRST} = C (\psi \bar{\psi} - J^3 - i\sqrt{n \over 2} \partial X).$$
The coefficient in front of $X$ is determined by requiring
the nilpotency of the BRST charge $Q_{BRST}$. This also implies that $X$
should live on $S^1$ of
radius $R=\sqrt{2/n}$. It follows that the $U(1)$ current $J$ in
\kazama\ is equal to \uone\ up to a BRST trivial operator.

Now we are ready to perform the orbifoldization of
the Landau-Ginzburg model, noting that the generator
of $G$ acting on the $SL(2)/U(1)$ piece can be identified with
${\rm exp}(2\pi i J_0)$ and so in particular using \uone\ it acts
only on the $X$ piece of the coset.
The relevant partition function of the $X$ system is given by
\eqn\suzuki{        {\vartheta_{m,n}(\tau) \over \eta(\tau)}  , }
where
$$ \vartheta_{m,n}(\tau) = \sum_{s \in {\bf Z}}
                  q^{n(s+{m \over 2n})^2}. $$
For the $SU(2)/U(1)$ system, which is the $N=2$ minimal model,
we use a character of an irreducible representation of the
$N=2$ superconformal algebra, which is denoted by
$ch_{l,m'}^{(n-2)}(\tau)$,
where $l$ and $m'$ are related to the conformal weight $h_{l,m'}$ and the
$U(1)$ charge of the highest weight state $Q_{m'}$ as
$$  h_{l,m'} = {l(l+2) - m'^2 \over 4n},~~
    Q_{m'} = {m' \over n} .  $$
The charge integrality condition, which
is imposed by the orbifoldization by $G$, requires that we choose
$m$ in \suzuki\ to be equal to $m'$ in $ch_{l,m'}^{(n-2)}$
so we only consider the combination
\eqn\product{       {\vartheta_{m,n}(\tau) \over \eta(\tau)} \cdot
  ch_{l,m}^{(n-2)}(\tau). }
To sum over the spectral flow, we make use of
the following expression for the $N=2$ character \ref\ry{
F.\ Ravanini and S.-K.\ Yang,``Modular invariance in $N=2$
superconformal field theories,'' {\it Phys. Lett.}
{\bf 195B} (1987) 202.}
$$ ch_{l,m}^{(n-2)}(\tau) = \sum_{m'=-n-1}^{n-2}
 c_{l,m'}^{(n-2)}(\tau) \vartheta_{m'n-m(n-2),n(n-2)}(\tau/2), $$
where $c_{l,m'}^{(n-2)}(\tau)$ is the string function ($=$
parafermionic partition function $ \times ~ \eta(\tau)$) of the affine
$SU(2)$ algebra. The string function can be obtained by
expanding a character of an irreducible
representation of the affine $SU(2)$ algebra at level $(n-2)$
with the highest weight $l$, $\chi_{l}^{(n-2)}(\tau)$,
in terms of the theta-function as
\eqn\string{ \chi^{(n-2)}_l(\tau) = \sum_{m}
       c_{l,m}^{(n-2)}(\tau) \vartheta_{m,n-2}(\tau). }
The information on the $U(1)$ charge is carried by
the theta-function piece of this expression. Thus the
problem reduces to computing
$$ \sum_{m} \vartheta_{m,n}(\tau)
        \vartheta_{m'n-m(n-2),n(n-2)}(\tau/2) .$$
This can be done by using the multiplication formula of the theta-function.
By using \string, we can recombine the string function
and the theta-function into the $SU(2)$ character as
\eqn\spectralflow{
\eqalign{ &
\sum_{m=-l}^{n-1-l}
     {\vartheta_{m,n}(\tau) \over \eta(\tau)} \cdot
  ch_{l,m}^{(n-2)}(\tau) = \cr
&= \prod_{r=1}^\infty (1+q^{r-1/2})^2
   \cdot \chi_{l}^{(n-2)}(\tau)  \cr} }
Thus we have shown that the orbifoldization
of the $N=2$ minimal model combined with
one free scalar $X$ coming from the
Kazama-Suzuki model gives the $SU(2)_{n-2}$ WZW for the $A$-series
(where the left and right characters are identical) plus two free fermions.
It is straight-forward to generalize this construction
for the $D$- and $E$-series of ALE above, with the modification that the
resulting $SU(2)$ partition function will be the $D$ or $E$-type modular
invariant \ref\ciz{A.\ Cappelli,
C.\ Itzykson and J.B.\ Zuber,
``The $ADE$ classification of minimal and $A_1^1$
conformal invariant theories,'' {\it Commun. Math. Phys.}
{\bf 113} (1987) 1.}.  This is clear because the corresponding
modular combinations for the $N=2$ minimal models
follows that of the underlying $SU(2)$.  Note that,
generally speaking, the orbifoldization of the $\hat{c}=2$ Landau-Ginzburg
model introduces the affine $SU(2)$ symmetry of level $1$, which
is a part of the (small) $N=4$ superconformal algebra \ref\bd{
T.\ Banks and L.\ Dixon, ``Constraints on string vacua with
space-time supersymmetry,'' {\it Nucl. Phys.} {\bf B307}(1988)
93.}, \ref\eoty{T.\ Eguchi, H.\ Ooguri, A.\ Taormina and S.-K.\ Yang,
``Superconformal algebras and string compactification on
manifolds with $SU(N)$ holonomy, {\it Nucl. Phys.} {\bf B315}(1989)
193.}, and which is different from the $SU(2)_{n-2}$ symmetry
manifesting itself in \spectralflow.

The remainder of the Kazama-Suzuki model, apart from the
two fermions of the $SL(2)/U(1)$ model, is
 a free scalar
$\phi$ with a background charge $Q_\phi \sim 1/\sqrt{n}$.
The easiest way to see it is to use the Wakimoto construction of
$SL(2,R)$ which realizes
$J^\pm ,J^3$ in terms of one scalar field $\phi$ and a pair of
bosonic ghosts $(\beta, \gamma)$ of weights $(1,0)$. Following
the analysis of \mv, one finds that $(\beta, \gamma)$ and $(B,C)$
makes the Kugo-Ojima quartet and decouple (in a suitable region
of field space `away' from the tip of the cigar). The remaining $\phi$
has a background charge $Q_\phi$, and the screening operators of the Wakimoto
construction go over to the screening operators for $\phi$.

To summarize, we have found that the orbifoldized Landau-Ginzburg
model with the superpotential given by \adesuperpotential\ is
equivalent to the $SU(2)_{n-2}$ WZW model, four free fermions and
one scalar with the background charge $Q_\phi$. In particular,
the resulting model has
$SU(2)_{n-2} \times SU(2)_1$ symmetry. Combined with
the free scalar $\phi$ and the (small) $N=4$ supercurrent,
they generate the (large) $N=4$ superconformal algebra \ref\sevrin{
A.\ Sevrin, W.\ Troost and A.\ Van Proeyen, ``Superconformal algebras
in two dimensions with $N=4$,'' {\it Phys. Lett.} {\bf 208B} 1988.}.

In claiming a connection between ALE spaces and LG models one
has to precisely specify the K\"ahler moduli in order
to get a singular conformal theory \aspin .  For example
consider the $A_1$ singularity.  If we consider it as
an orbifold $R^4/Z_2$, the usual orbifold CFT is not singular
precisely because the periodic $B$ field in the orbifold
model is $B=\pi$ and not zero.  If we adjust $B$, keeping
the sphere collapsed, there are two values for which the resulting theory has
an extra $Z_2$ symmetry: $B=\pi,0$.  The case $B=0$ is the one which
gives the singular conformal theory.
In fact, in our models, the
K\"ahler moduli are automatically tuned such that the $B$ fields are zero.
To see this let us consider the $A_1$ case as an example.
In general, the orbifoldized
Laudau-Ginzburg models have discrete symmetries so called
quantum symmetries, which
fix the value of the the K\"ahler moduli \ref\vafasym{
C.\ Vafa, ``Quantum symmetries of string vacua,''
{\it Mod. Phys. Lett.} {\bf A4} (1989) 1615.}.  For the
case of $A_1$ considered above we have to mod out by a $Z_2$
so we will indeed get a $Z_2$ quantum symmetry in the orbifoldized model.
So we are either at $B=0$ or $B=\pi$.  Since the concrete
description of our conformal theory given above shows that
it is not equivalent to standard $R^4/Z_2$ orbifold CFT, which
has $B=\pi$ we conclude that $B=0$ in the above construction.

As mentioned before for the case of the $A_{n-1}$ series, the orbifold
Laudau-Ginzburg model we just constructed is identical to
the conformal field theory of the symmetric
fivebrane solution with the
$H$ charge $n$ \ref\hs{G.\ Horowitz and
A.\ Strominger, ``Black strings and $p$-branes,'' {\it Nucl. Phys.}
{\bf B360} (1991) 197.}, \chs,
\ref\chstrieste{C.G.\ Callan, J.A.\ Harvey and A.\ Strominger,
``Supersymmetric string solitons,'' in Proceeding of
the 1991 Trieste Spring School and String Theory and Quantum Gravity,
eds. }. The target space geometry of the fivebrane solution has
an asymptotically flat region and there is a semi-infinite cylinder
(wormhole) attached to its center.
The scalar field $\phi$  corresponds to the radial coordinate along
the cylinder, and the $SU(2)_{n-2}$ WZW model describes the $S^3$
part, which is the cross-section of the cylinder. The volume of the
cross-section is proportional to the $H$ charge $n$ \hs.
Thus if this were a bosonic system, the level of the $SU(2)$ algebra would
be equal to $n$. However we must also take into account four fermions which
are superpartners of $\phi$ and the $SU(2)$ WZW model. Initially they are
are not free but are coupled to the generalized connections on the
target space. However in the fivebrane solution, these connections
are parallelized by the anti-symmetric tensor field and we can transform
the fermions into free ones by chiral gauge rotations. This procedure causes
the shift $n \rightarrow n-2$ through chiral anomalies.

Recently Witten has examined the phase structure of
the linear sigma model for the $A_1$ type singularity
and conjectured that it should be related to the fivebrane solution
based on the similarity between their global geometric structures
\witju. Here we have proven
his conjecture by studying the Hilbert space structure of the model
explicitly, and we have determined the value of the $H$ charge to be $2$.
We also extended it to all $A_{n-1}$, with the
result that the ALE singularity given by $A_{n-1}$ is equivalent
to the conformal theory describing the symmetric fivebrane with the
charge $n$.
There are two subtleties to consider:  The first one is that if we consider
type IIA(B) on ALE it is equivalent to type IIB(A) on the symmetric fivebrane.
This exchange is easy to see, because in the LG model above we
could vary the complex structure of ALE, i.e. it is a B-model
description of the theory, whereas in the symmetric fivebrane, the
description is in the A-model (i.e. the sigma model origin for supercharges).
This statement will become even more clear in the next section which
is related to resolving yet another puzzle:  How come we got the $H$
field and the dilaton field turned on, while we thought we were talking
about $K3$ target space?  A hyperk\"ahler metric on $K3$ is
uncorrected in worldsheet theory and gives rise to a conformal
theory.   Where do we get the dilaton and $H$ fields from?
This is the topic of the next section, where we use the stringy
cosmic string description of $K3$ given in \scs\ to show
how the $H$ and dilaton arises as well as rederiving the
relation between the value of charge of the symmetric fivebrane and
the value of $n$ in the $A_{n-1}$ type singularity.

\newsec{ Stringy Cosmic Strings, $K3$ and its Dual}
Let us recall the construction of stringy cosmic string \scs .
We start with compactification on a two torus $T^2$.  The moduli
for this compactification is given by two complex parameters $(\tau,\rho )$
where $\tau$ parameterizes the complex structure of $T^2$ and $\rho$
parameterizes the complexified K\"ahler class of $T^2$.
Note that both are defined up to an $SL(2,{\bf Z})$ transformation.
Moreover there is an exchange symmetry $\tau \leftrightarrow \rho$
which is obtained by applying $R\rightarrow 1/R$ to one of the circles.
 Viewed
in eight-dimensional terms, the moduli $\tau$ and $\rho$ appear
as complex scalar fields.  We consider finding
solutions to the low energy eight-dimensional equations of motion, where fields
depend only on 2 extra parameters (i.e. we are interested
in keeping a six-dimensional Poincare invariance).
Let us denote the extra two-dimensional space by $z$.  A nice class of such
solutions \scs\ can be obtained by taking
$\tau (z)$ to
depend holomorphically on $z$, and to take $\rho$ to be constant,
independent of $z$.  In making sense of such a solution one
has to recall that $\tau$ is defined only up to $SL(2,{\bf Z})$
and so we are allowed to have jumps in $\tau$ up to this transformation.
Geometrically what this means is that as we go around some cycles
on the $z$-plane, if we follow the compactified $T^2$, it comes back
to itself, but with some monodromy.  In other words if we follow a
given cycle of $T^2$ continuously it does not necessarily come back to itself.
Geometrically this is a perfectly nice condition. If around some point say
$z=0$ we have monodromies, it must mean that the $T^2$ fiber
above it is singular where a cycle vanishes, and the monodromy
of cycles around $z=0$ can give additions of this vanishing cycles.
Recall that a degenerating $T^2$ can be viewed near its degeneration
point by
$$xy=q$$
where $x,y$ are coordinates on some patch, and $q$ denotes the plumbing
fixture
modulus.  As $q\rightarrow 0$ we get a torus with a vanishing cycle.
Now let us replace $q\rightarrow z$ and think of the degenerating family
of elliptic curves over the $z$-plane given locally by $xy=z$.  At
$z=0$ we have our degenerating torus.  Note however, that the total space
is {\it not} singular despite the fact that the fiber is singular.
The easiest way to see this is to consider $x,y$ as the coordinates.
Clearly $z$ is fixed uniquely in terms of $x,y$ as $z=xy$ and so the
space looks locally like ${\bf C}^2$.
The points over which the fiber degenerates are the points where
there is a `cosmic string' (if we were dealing with compactifications
down to 4).

We can, however, develop a singularity if we get $n>1$ singular
fibers approach each other.  Let $a_i$ denote the $z$ coordinate
of $n$ singular fibers with $i=1,...,n$.  Suppose the singularities
are such that the point on $T^2$ where they degenerate are the same.
Then a local description of this singularity can be given as
$$xy=\prod_i (z-a_i)$$
and if we take $a_i\rightarrow 0$ we would be discussing the $A_{n-1}$
singularity at $x=y=z=0$.

In solving the eight-dimensional
equations of motion, one realizes that since there is energy density
due to the variation of the scalar field $\tau$ there is curvature.
In fact it was shown in \scs\ that if we have $n$ points where the fiber
degenerates in the way we discussed above the $z$-plane will have a
conical asymptotic with conical angle given by $2\pi(1 -{n\over 12})$.
In particular with $n=12$ we get a cylindrical behavior and with
$n=24$ the $z$-plane compactifies to a sphere and the total space is
compact.  This compact space is nothing but $K3$!

The low energy equations of motion that were solved to give
the stringy cosmic string solution are inapplicable precisely
where the fiber degenerates.  However it was argued in \scs\
that there exists a metric on the total space which leads to
a hyperK\"ahler metric on the total space. This is also clear in the
case where $n=24$ as $K3$ admits a hyperK\"ahler metric and we
can view $K3$ as an elliptic fibration over the sphere (i.e.
$K3$ is an elliptic surface).

It was also observed in \scs\ that we can exchange the role
of $\tau \leftrightarrow \rho$ in the above construction,
and the $T$-duality of the problem implies that we must have
still a solution to the equations of motion. This of course means
that type IIA in one case is equivalent to type IIB on the dual.
Once we apply the duality,
the metric of $T^2$ is now varying as well as the $B$ field, because
$\rho ={B\over 2\pi}+i\sqrt G$ on $T^2$.  Thus the analog
of the singular fibers around which we
have the monodromy $\tau \rightarrow \tau+1$ now corresponds
to a fixed $\tau$ with $\rho \rightarrow \rho+1$, i.e.,
$B\rightarrow B+2\pi$.  This in particular means that if we take
a three cycle consisting of a circle $S^1$ around the point with
degenerate fiber, times the fiber $T^2$ and integrate
$${1\over 2\pi}\int_{T^2\times S^1}H=\int_{T^2\times S^1}
{dB\over 2\pi}={1\over 2\pi}\Delta B\big|_{S^1}=1$$
we have one unit of $H$ charge in that region. Note also that hidden in this
discussion is that there is now also a dilaton
turned on.  To see this note that the duality that exchanges $\tau
\leftrightarrow \rho$ also changes the dilaton so that
$\rho_2/\lambda^2 =\rho_2 {\rm exp }(-2\phi)$
is invariant.  Thus in the new solution $
{\rm exp }(2\phi)=A \sqrt G$ for some constant $A$.  Note that
if we consider the case where $n$ cosmic strings come together,
where we would get $A_{n-1}$ singularity,
we will have in the dual $\tau \leftrightarrow \rho$
description $n$ units of $H$ charge, with the dilaton turned.
This is beginning
to look like the symmetric fivebrane solution of \chs.  However
to be more explicit we need to go beyond the adiabatic approximation
used in \scs\ where roughly speaking the metric is split to a block diagonal
form involving $T^2\times S^2$.  Given that the singularity takes
place at a given point in four space, this adiabatic approximation,
which treats all points on $T^2$ symmetrically,
misses even the symmetries of the leading singularity.
Note that the singularity occurs at a given point in
four space, which is of the same geometry as the
symmetric fivebrane and given that it
carries an $H$ charge, it is natural to expect that the
leading singular piece of the
two conformal theories be equivalent,
as was demonstrated in the previous section.

\newsec{$K3$ and Calabi-Yau Threefold Singularities}
In section three of this paper we discussed the singularity
corresponding to ALE spaces and in section two that of the conifold
singularity.  Is there a connection between them?  The answer is
yes, as anticipated in the description of conifold singularity
given in \ref\hub{T.\ H\"ubsch, ``Calabi-Yau manifolds, a bestiary for
physicists,'' (World Scientific, 1992). }\
and used in \dstring :  We can consider
compactifying down to six dimensions on $T^2\times T^2$ and talk
about the variation of each of the complex structures $\tau_1$
and $\tau_2$ of the two
$T^2$ as a function of an extra $z$-plane (this was also briefly
considered in \scs ).  Now as a function of the $z$-plane
the $\tau_1(z)$ and $\tau_2(z)$ might be singular at different
points.  Let us denote the coordinates of the degenerating
first torus by `plumbing fixture' coordinates $x,y$ as before,
and those of the second one by $u,v$.  Let us assume that the
first torus degenerates at $z=0$ and the second one at $z=\mu$
for $\mu $ close to zero.  Then locally the description of the
three manifold looks as
$$xy=z \qquad uv=z-\mu$$
which, by eliminating $z$ is equivalent to
$$xy-uv=\mu$$
which is the conifold singularity.  We will now describe
this in the conformal theory language and show how the conifold
conformal theory is related to that for the ALE space.
Before doing so, let us generalize the above singularity
to the one corresponding to $c=1$ at radius $n$ times the self-dual value.
In that case the singularity can still be given a description as above
where $n$ cosmic strings of the first torus come together on the $z$-plane,
i.e. we have
\eqn\gener{xy=z^n \qquad uv=z-\mu}
which upon eliminating the $z$ variable give the $R=n/\sqrt 2$
singularity \ref\mukgo{D. Ghoshal, P. Lakdawala and S. Mukhi,
Mod. Phys. Lett. {\bf A8} (1993) 3187.}
$$xy=(uv+\mu)^n$$
Now it is clear how to write down the conformal theory
corresponding to these singularities.  Consider the Landau-Ginzburg
theory
$$W=\zeta^{-n}+z^n+a^2+b^2 +\Lambda (z-xy).$$
This is the same superpotential as the $A_{n-1}$ singularity
except for the additional fields $\Lambda$,$x$ and $y$ and the
interaction $\Lambda (z-xy)$.  Upon integrating out $\Lambda$
we would get the identification of $z=xy$ and the superpotential
becomes
$$W=\zeta^{-n}+(xy)^n+a^2+b^2$$
Note that for $n=1$
this is the same superconformal theory
as we discussed in section two, for the Calabi-Yau conifold
and it is known.  For other values of $n$, apart from
the above LG description, we do not have the full conformal
theory description.  Anyhow this rederivation of the conifold
conformal theory shows a parallel consistency between the two classes
of conformal theories we discussed and their geometric relation
between the two explained above.  It also explains the appearance of
two-dimensional
 Euclidean black hole in the discussion of the conifold singularity.

This also raises the question of the conformal theory of `one
cosmic string', i.e. symmetric fivebrane with charge $1$.  This
is the case where the level of the $SU(2)$ theory is $-1$, which presumably
means that we have to interpret it as $SL(2)$ at level $1$.  Thus
we end up with the product of two $SL(2)/U(1)$'s one at level 1 and
the other at level 3.  So
it seems that the geometry for one symmetric fivebrane has a different
description from the $n>1$ case.  It would be worthwhile studying
this further.

\newsec{$K3$ Singularity and $c < 1$ Topological String}

In section two, we have seen that topological string amplitudes
on a Calabi-Yau manifold near the conifold are computable by
transforming them into the $c=1$ string amplitudes. Here we
will provide
some evidence that topological string amplitudes on $K3$ near
the $ADE$ singularities are related to the $c<1$ topological
strings \ref\keke{K.\ Li, ``Topological gravity with minimal matter,''
{\it Nucl. Phys.} {\bf B354} (1991) 711; ``Recursion relations
in topological gravity with minimal matter,'' {\it Nucl. Phys.}
{\bf B354} 725.},
\ref\dw{R.\ Dijkgraaf and E.\ Witten, ``Mean field theory,
topological field theory, and multimatrix models,''
{\it Nucl. Phys.} {\bf B342} (1990) 486.},
\ref\dvv{R.\ Dijkgraaf, E.\ Verlinde and H.\ Verlinde,
``Topological strings
in $d < 1$, {\it Nucl. Phys.} {\bf B352} (1991) 59.}\
corresponding
to the $ADE$ $(1,n)$ matter coupled to gravity.

The first evidence for this relation is that the
$y$-part of the superpotential \adesuperpotential\
is that for the $N=2$ minimal model, whose topological twisting gives the
standard description of the $c<1$ topological string \keke.
The relation between the $c<1$ topological string
and the twisted minimal model is further clarified in
\ref\blnw{B.\ Gato-Riviera and A.M.\ Semikhatov,
``Minimal models from $W$ constrained hierarchies via
the Kontsevich-Miwa transform,'' {\it Phys. Lett.}
{\bf B288} (1992) 38;
M.\ Bershadsky, W.\ Lerche, D.\ Nemeshansky and
N.P.\ Warner, ``Extended $N=2$ superconformal structure of gravity
and $W$ gravity coupled to matter,'' {\it Nucl. Phys.}
{\bf B401} (1993) 304.} where it is pointed out that
the system of the Liouville field, the $(1,n)$ matter and
the BRST ghosts has the $N=2$ superconformal symmetry
of $\hat{c} = 1 - 2/n$ with the same chiral ring structure
as that of the twisted $N=2$ minimal model of the $A_{n-1}$ type.
However there is another choice of twist (by using the matter
current instead of the Liouville)
 which leads to an $N=2$ algebra
with $\hat{c} = 1 + 2/n$. This observation, in the case of
$n=1$, led to the
description of the $c=1$ string at the self-dual radius
in terms of the Landau-Ginzburg model with $W = x^{-1}$ \mv.
Therefore the $c<1$ string may also be related to the model with
$W = x^{-n}$, in the case of the $A_{n-1}$ series.
Curiously the superpotential \adesuperpotential\
for the $ADE$ singularity is combining these two descriptions
of the $c<1$ string.

The relation between the $W=x^{-n}$ potential and the $c<1$ string
can be stated more explicitly.
Let us first examine the case of $n=2$ by using
the description in term of the Kazama-Suzuki model
$SL(2,R)_4/U(1)$. In this case, the $U(1)$ current of the
$N=2$ algebra \uone\ takes the form
$$  J = 2\psi\bar{\psi} -J^3  $$
and the BRST current is
$$ J_{BRST} = C(\psi\bar{\psi} - J^3 -i \partial X) .$$
In this case, the scalar $X$ lives on $S^1$ with $R = 1$.
We can then rewrite the $U(1)$ current $J$
as
$$ J = 3\psi\bar{\psi} - 2J^3 - i \partial X
            - \{ Q_{BRST}, B \} . $$
After the topological twisting by $J$, $(\bar{\psi}, \psi)$ become
of weights $(2, -1)$ and we may identify them as the BRST ghosts
for diffeomorphism. Since $X$ is at radius $R=1$, we can fermionize
it to $(b, c)$ which become of weights $(1,0)$ after the twisting.
As in the case of the $c=1$ string, $(\beta,\gamma)$
of the Wakimoto construction of $SL(2,R)_4$ decouple together with
$(B,C)$, and we are left with the scalar field $\phi$ with
$c = 27$. We can then combine $(b, c)$
and $\phi$ to make bosonic ghosts $(\tilde{\beta}, \tilde{\gamma})$
of weights $(2,-1)$. The resulting system of $(\bar{\psi},\psi)$ and
$(\tilde{\beta}, \tilde{\gamma})$ is nothing but the pure topological
gravity  \ref\witgr{E.\ Witten, ``On the structure of
the topological phase of two-dimensional gravity,''
{\it Nucl. Phys.} {\bf B340} 281.},
\ref\distler{J.\ Distler, ``2-d quantum gravity, topological field
theory and multicritical matrix models,'' {\it Nucl. Phys.}
{\bf B342} 523.}, \ref\vv{
E.\ Verlinde and H.\ Verlinde, ``A solution of two-dimensional topological
quantum gravity,'' {\it Nucl. Phys.} {\bf B348} 457.}.

For general $n$, the $U(1)$ current \uone\ of the Kazama-Suzuki
model can be written as
$$ J = 3\partial\bar{\psi} - 2 J^3 - i\sqrt{2 \over n} (n-1) \partial X
             - 2(1 -{1 \over n}) \{ Q_{BRST}, B \}. $$
Therefore the topological twist by  $J$ again transforms
$(\psi,\bar{\psi})$ into the BRST ghosts of weights $(2,-1)$.
The central charge for the scalar $\phi$ becomes
$$  c_\phi = 1 + 6 {(n+1)^2 \over n}  , $$
and the scalar $X$ has
$$ c_X = 1 - 6 {(n-1)^2 \over n} . $$
In other words the topological theory we are dealing is equivalent
to the $(1,n)$ matter coupled to the two-dimensional gravity.

Thus we are led to conjecture that the topological string theory
on the $ADE$ singularities \ade\ is related to
the $c<1$ topological string
\keke, \dw, \dvv. In the case of $n>2$, however, there is
a subtlety since we need to
perform the orbifoldization of the Landau-Ginzburg model
as in section two and therefore the $SL(2,R)_{n+2}/U(1)$
Kazama-Suzuki model and the $N=2$ minimal model are not completely
decoupled. To clarify this point, let us work on the
Landau-Ginzburg model directly.

For the $\hat{c}=2$ model, the most natural class of string
amplitudes are $N=2$ string amplitudes \ref\ov{H. Ooguri and C. Vafa,
``Selfduality and $N=2$ String Magic,''
{\it Mod. Phys. Lett.} {\bf A5} (1990) 1389;
H. Ooguri and C. Vafa, ``Geometry of $N=2$ Strings,''
{\it Nucl. Phys.} {\bf B361} (1991) 469.}, which can be expressed
as $N=4$ topological string amplitudes \bv. Let us consider
the $A_{n-1}$ singularity described by the superpotential
$$ W = - \mu x^{-n} + y^{n}. $$
This potential can be deformed by adding chiral primary
fields. In order to preserve the $N=4$ symmetry, we only
allow chiral primary fields of charge $1$ given by
\eqn\chiral{ \phi_i = x^{i-n-1} y^{i-1}  ~~~(i=1,2,...,n-1). }
In particular, $\phi_{i=1}$ changes the value of $\mu$.
At genus-$0$, $N=4$ topological amplitudes are defined
for four or higher points \bv\ as
\eqn\neqfour{
\eqalign{ A_{i_1 \cdots i_m}^0 =
  \int d^2z_4 \cdots d^2 z_n & \langle
  \phi_{i_1}(0) \phi_{i_2}(1) \phi_{i_3}(\infty) \times \cr
 &~ \times J^{-}_L J^{-}_R \phi_{i_4} (z_4)
  G^-_L G^-_R \phi_{i_5}(z_5) \cdots
  G^-_L G^-_R \phi_{i_n}(z_n)   \rangle , \cr}}
where $J^-$ is one of the $SU(2)$ current of level $1$
which belongs to the (small) $N=4$ algebra.

It turns out that the selection rule for this amplitude
is consistent with that of the $(1,n+1)$ matter
coupled to the two-dimensional
 gravity. The $(1,n+1)$ matter can be realized
as a twisted $N=2$ minimal model \keke\ with the superpotential
$$  W(y) = y^{n+1} . $$
This superpotential can be deformed by adding chiral primary
fields given by
$$  \phi'_i = y^i ~~~(i=0,1,...,n-1) $$
If we consider four or higher points at genus-$0$, a topological
string amplitude involving $\phi'_{i=0}$ vanishes since
$G^-$ annihilates it. Therefore only $\phi'_i$ with
$(i=1,...,n)$ are relevant in the comparison.
By using the $U(1)$ charge conservation
and the parafermion selection rule, it is straightforward to
show that the selection rule for the
$(1,n+1)$ topological string amplitude
\keke, \dw, \dvv\ satisfies the selection rule
for the $N=4$ string amplitude \neqfour\
provided we identify $\phi_i'$ with $\phi_i$
($i=1, 2,...,n-1$).  It is not yet clear why the shift
$n \rightarrow n+1$ of the level takes place, and this
merits further investigation.

\bigskip
We would like to thank S.\ Shenker, who participated in an earlier
stage of this project, for discussions.
We also thank the Aspen Center for Physics, where this project was
initiated. The work of HO was supported in part by the National
Science Foundation under grant PHY-9501018 and in part by
the Director, Office of Energy Research, Office of High Energy
and Nuclear Physics, Division of High Energy Physics of the
U.S. Department of Energy under Contract DE-AC03-76SF00098.
The research of CV was supported in part by NSF grant PHY-92-18167.

\listrefs

\end